\documentclass[aps,prd,twocolumn,
showpacs,amssymb,
nofootinbib]{revtex4}%twocolumn,

\usepackage{amssymb,amsmath}
\usepackage{amsfonts}
\usepackage{mathrsfs}
\usepackage{txfonts}
\usepackage{marvosym}
\usepackage{amssymb}

\begin{document}

%%%%%%%%%%%%%%%%%%%%%%%%%%%%%%%%%%%%%%%%%%%%%%%%%%
%%%%%%%%%%%%%  TITLE   %%%%%%%%%%%%%%%%%%%%%%%%%%%

\title{Uniqueness of the Fock quantization of fields with unitary
dynamics in nonstationary spacetimes}

\author{Jer\'onimo Cortez} \email{jacq@fciencias.unam.mx}
\affiliation{Departamento de F\'\i sica, Facultad de
Ciencias, Universidad Nacional Aut\'onoma de M\'exico,
M\'exico D.F. 04510, Mexico.}

\author{Guillermo A. Mena
Marug\'an}\email{mena@iem.cfmac.csic.es} \author{Javier
Olmedo}\email{olmedo@iem.cfmac.csic.es}
\affiliation{Instituto de Estructura de la Materia, CSIC,
Serrano 121, 28006 Madrid, Spain.}

\author{Jos\'e M. Velhinho}\email{jvelhi@ubi.pt}
\affiliation{Departamento de F\'{\i}sica, Universidade da
Beira Interior, R. Marqu\^es D'\'Avila e Bolama, 6201-001
Covilh\~a, Portugal.}

\begin{abstract}

The Fock quantization of fields propagating in
cosmological spacetimes is not uniquely determined
because of several
reasons. Apart from the ambiguity in the choice of
the quantum representation of the canonical commutation
relations, there also exists certain freedom in the choice of
field: one can scale it arbitrarily absorbing background
functions, which are spatially homogeneous but depend on time.
Each nontrivial scaling turns out into a
different dynamics and,
in general, into an inequivalent
quantum field theory. In this work we analyze this freedom
at the quantum level for a scalar field in a nonstationary,
homogeneous spacetime whose spatial sections have $S^3$
topology. A scaling of the configuration variable is introduced as
part of a linear, time dependent canonical
transformation in phase space. In this context, we prove in full
detail a uniqueness result about the Fock quantization
requiring that the dynamics be unitary and the spatial
symmetries of the field equations have a natural unitary implementation.
The main conclusion is that, with those requirements,
only one particular canonical transformation is allowed, and
thus only one choice of field-momentum pair (up to irrelevant constant
scalings).
This complements another previous uniqueness result for scalar
fields with a time varying mass on $S^3$, which selects
a specific
equivalence class of Fock representations of the canonical commutation
relations under the
conditions of a unitary evolution and the invariance of the vacuum
under the background symmetries. In total,
the combination of these two different statements of uniqueness
picks up a unique Fock quantization for the system.
We also extend our proof of uniqueness
to other compact topologies and spacetime dimensions.

\end{abstract}


\maketitle

\section{Introduction}
\label{sec:intro}

The unique character of Nature is alluded in physics
by the uniqueness of the theories employed to describe
it. In particular, by imposing appropriate physical
criteria, the quantization of a classical system
{\sl{should}} yield a unique quantum description --up
to unitary equivalence. Since the quantization process
involves choices that may lead to inequivalent
theories, the specification of a unique description is
a nontrivial task.

Even in systems in which one already starts with a {\sl specific}
choice of basic canonical variables and an associated set of
canonical commutation relations (CCR's), there exists an intrinsic
ambiguity in the quantization process because the CCR's can be
represented in nonequivalent ways. In the case of linear
systems with a finite number of degrees of freedom,
these ambiguities are essentially suppressed by
the imposition of certain unitarity and
continuity conditions on the representation of the
algebra of observables (as stated in the Stone-von Newmann
theorem \cite{Vonneu}), so that uniqueness follows
without misadventures. Nonetheless, the situation
changes drastically in the arena of field theory.
These systems accept infinite nonequivalent
representations of the CCR's \cite{wald} and
there is no general procedure
to select a preferred quantum description. In this
situation, {\sl physical results} depend on the
representation adopted, a fact that brings into
question their significance. It is then necessary to
look for additional criteria to warrant
uniqueness and regain robustness in the quantum
predictions.

The usual
procedure to select a preferred
representation in field theory for a {\sl given} set of CCR's
is to exploit the
classical symmetries. For instance, the invariance
under the Poincar\'e group is the criterion imposed to
arrive at a unique representation in ordinary quantum
field theory. Thus, if the field theory corresponds to
a scalar field, Poincar\'e invariance,
adapted to the
dynamics of the considered theory, selects a
complex structure \cite{wald}, which is the
mathematical object that encodes the ambiguity in the
quantization and determines the vacuum state of the
Fock representation. For stationary spacetimes, the
time translation symmetry is exploited to formulate
the so-called energy criterion and then select a
preferred complex structure \cite{a-m}. But when the
symmetries are severely restricted, as it is the case
for generic curved spacetimes or for manifestly
nonstationary systems, extra requirements must be imposed to
complete the quantization process. For example, in
the case of de Sitter space in $1+1$ dimensions,
it is possible to pick up a unique de Sitter
invariant Fock vacuum for a free scalar field
by looking for an invariant Gaussian solution to a properly
regulated Schr\"odinger equation \cite{jackiw}.

In the context of quantum cosmology, the
extra criterion of a unitary implementation of the
dynamics has been successfully employed to specify a
unique, preferred Fock quantization for the Gowdy
spacetimes. These are spacetimes
which possess two spacelike Killing isometries
and spatial sections of compact topology \cite{gowdy}.
In the case of a three-torus topology and a
content of linearly polarized gravitational waves,
the local gravitational degrees of
freedom can be described by a scalar field with a
specific time dependent mass and which propagates in
an auxiliary, static background with the spatial
topology of the circle \cite{unit-gt3}.
For this choice of basic
field for the model, one is able to find a
unique Fock quantization which
incorporates the background symmetries as
symmetries of the vacuum and
implements the field dynamics
as a family of unitary quantum transformations
\cite{feasab,unit-gt3,unique-gowdy-1,CMV5}.

More recently, in a broader context,
a unitary equivalence class of Fock representations has been specified for scalar
fields with {\sl generic} time varying mass, defined
on spheres in three or less dimensions \cite{cmsv,CMV8}. Again, the
procedure consists in requiring a unitary dynamics and the
vacuum invariance under the symmetries of the field equation.
The particularly relevant case of the three-sphere, with the
dimensionality observed in our universe, was considered
in Ref. \cite{CMV8}.

Apart from the inherent ambiguity in choosing the representation
of the CCR's,
the quantization of fields in curved space-times
is affected by another kind of ambiguity. It is due to
the freedom in choosing a specific field {\sl parametrization} to describe
the physical system, namely, the freedom in declaring a particular choice
of field (together with its associated dynamics)
as the {\sl fundamental} one. Let us concentrate our attention
on the case of homogeneous but nonstationary spacetimes, like those
encountered as backgrounds in cosmology. In these circumstances, it is most
natural to consider field redefinitions which absorb background
functions.
This leads to a scaling of the field by a time dependent
function, such that the linearity of the field equations and of the structures
of the system are preserved. If this time dependence is nontrivial, the two fields
(i.e., the scaled and the unscaled ones) are governed by different
dynamics. Since a change in the dynamics typically calls for inequivalent
representations, the construction of a quantum theory clearly depends
on the selection of a specific field description for the system among all those related
by these scaling transformations.

As commented above, these considerations are crucial for quantum matter
fields propagating in inflationary or cosmological backgrounds,
which are spatially homogeneous but not stationary. The discussion is also
relevant for the quantization of local gravitational degrees of freedom,
in contrast with the previous context of quantum matter fields in classical
spacetimes that are solutions to the gravitational field equations.
This latter class of systems includes, e.g., the already mentioned Gowdy
models and the case of gravitational perturbations around cosmological
backgrounds. For these gravitational systems, there exists a great freedom
in the choice of parametrization of the metric components in terms of
fields. In all these situations, the choice of a suitable field
parametrization involves a time dependent scaling related
to background functions and whose specific form
depends on the particular system under study. This
choice often leads to fields which effectively propagate in an auxiliary
static background, therefore simplifying in part the corresponding
dynamics, although there remain (or appear) time dependent
potentials which manifest that the scenario is a nonstationary one.

The question immediately arises of whether
it is again possible
to invoke natural criteria to
remove (at least in certain situations) the ambiguity that this freedom in
the choice of field introduces at the quantum level.
A detailed analysis about this issue
was first carried out for the quantization
of the linearly polarized Gowdy model with three-torus topology in Ref.
\cite{CMV5}.
That work studied a family of linear,
time dependent canonical transformations that involve a scaling
of the field. It was proven that there actually exists no freedom
left in performing a transformation of this kind,
once the criteria of invariance under the remaining spatial symmetries
and the unitary implementation of the dynamics are
imposed. More precisely, Ref.
\cite{CMV5} shows that the considered transformations
lead to new dynamics such that one cannot attain a unitary quantum
evolution in a Fock representation while keeping the symmetry
invariance of the vacuum.\footnote{In fact, in this particular linear system
one can still introduce a redefinition
of the momentum which implies no scaling of the field,
but this turns out to be irrelevant inasmuch as
no new nonequivalent Fock quantization arises.}
The requirements of unitary evolution and invariance therefore
suffice to select a specific scaling of the field and a privileged
family of equivalent Fock quantizations for it. In other words, the
uniqueness is guaranteed both for the choice of fundamental field
(with its corresponding dynamics) and
for the quantum representation of the corresponding CCR's.

One may wonder whether
the uniqueness in the choice of field description can also be guaranteed
in other, more general systems than the Gowdy model,
and in particular for nonstationary settings where
there already exist results about the uniqueness
of the representation of the CCR's. The case of fields in
1+3 dimensional spacetimes with compact spatial topology is specially
important, owing to its applications e.g. to cosmology. A summary of the
discussion for scalar fields propagating in a nonstationary spacetime
with sections of $S^3$ topology was already presented by us in
Ref. \cite{CMOV}, anticipating that the answer to the question of
uniqueness is in the affirmative. The aim of the present work is to
provide full details of the demonstration of this result.

We will consider a scaling of the field by a generic function of time.
This scaling can always be completed into a time
dependent canonical transformation. We demand such transformation to
be compatible with all linear structures on phase space and with the
symmetries of the field equations. Any admissible canonical
transformation is then linear and, furthermore, can
be divided into two parts. The first one
is a linear canonical transformation that is explicitly time independent
but takes into account the initial conditions, rendering simple ones
for the remaining part, which incorporates then all the time
dependence.
We will demonstrate that there exists only one possible choice of
phase space variables such that the resulting field theory admits
a Fock quantization with unitary dynamics and a natural
implementation of the symmetries of the field equations.
The unique choice which remains available is precisely the one
which corresponds to a transformed scalar field that propagates
in a static spacetime with $S^3$ spatial topology,
though in the presence of a time varying mass term. Recall
that, for this latter field, the uniqueness
of the representation of the CCR's was proven in Ref. \cite{CMV8}.

As we have already mentioned, the list of scenarios where this result
finds direct applications includes the case of
inflationary models where a scalar field with constant mass
propagates in a Friedmann-Robertson-Walker (FRW) spacetime with
compact spatial topology. In this case, one can check that a linear,
time
dependent canonical transformation allows one
to rewrite the field equation as that of a field in a spacetime
with identical spatial topology but static, whereas the mass becomes
time varying. Another type of situations where
our result has implications is given by the quantization of
(inhomogeneous) perturbations around nonstationary
homogeneous solutions of the Einstein equations, typically cosmological
backgrounds \cite{lif,bardeen,bar2,mukhanov,mukhanov1}.
Examples are the gauge-invariant energy density perturbation
amplitude in an FRW spacetime with $S^3$ spatial topology
filled with a perfect fluid (when the perturbations of
the energy-momentum tensor are adiabatic
\cite{bardeen,mukhanov}) or the matter perturbations around
the same FRW spacetime for a massive scalar field \cite{hall_haw}.
With a suitable scaling (and in an appropriate gauge
in the case of the massive field),
the corresponding equations of motion can be related to
those of a scalar field in a static spacetime with a time
dependent mass term (see Ref. \cite{CMV5} for additional details).
At this point, it may be worth commenting that, although flat FRW 
universes receive a special attention in 
cosmology nowadays, some recent works find reasons to prefer 
closed FRW models with $S^3$ topology, for example from the point 
of view of perturbation theory in relation with the choice of 
appropriate gauges which embody Mach's principle \cite{bkl}, or 
in an attempt to account for a low microwave background 
quadrupole \cite{quad1,quad2}. On the other hand, we will
argue later on that our results can be 
generalized to the case of flat but compact FRW universes.

In summary, the question that we are going to investigate is whether
the criteria of unitary dynamics and invariance of the vacuum under
the symmetries of the field equation
select a unique Fock quantization among all those
arising from different time dependent scalings of the field.
We will concentrate our discussion on the case that
a particular scaling renders the dynamics into that of
a scalar field with time varying mass propagating
in a static spacetime, with inertial spatial sections that have
the topology of a three-sphere. We will study at the quantum level
the consequences of local, time dependent canonical transformations
which involve a scaling of the field. These transformations
must preserve the invariance of the field equation under the group of
symmetries and the linearity of the space of solutions.
Transformations of this kind consist
of a scaling of the configuration variable by a function
of time, the inverse scaling of the canonical momentum, and
possibly a contribution to the momentum that is linear in
the configuration variable, the proportionality
factor being time dependent. Our main goal in this work is to
provide a full proof demonstrating that the canonical
transformation is so severely restricted by our criteria
that it turns out to be fixed. In addition,
we will argue that the analysis can be
generalized to lower dimensions, replacing the three-sphere with
$S^2$ or $S^1$ (for this last case, see Ref. \cite{CMV5}), as well
as to other compact topologies.

The content of the paper is organized as follows. In
Sec. \ref{sec:static} we summarize the results that are already known
about the Fock
quantization of a scalar field with a time varying mass in
a static spacetime whose spatial sections have the
topology of $S^3$. In Sec.
\ref{sec:trans} we introduce the linear,
time dependent canonical
transformation which accounts for the scaling of the
field and discuss its consequences at the quantum level.
Sec. \ref{sec:uniq} contains the detailed proof that
only one of these canonical transformations
leads to a field dynamics which is compatible with our criteria of
a quantum unitary evolution and the symmetry invariance of the vacuum.
We discuss the results and conclude in Sec. \ref{sec:conc}.
Finally, an appendix which deals with some technical parts of our
demonstration is added.

\section{Preliminaries: The system and the Fock quantization of reference}
\label{sec:static}

Let us start by reviewing some of the key aspects and
results about the Fock quantization of a real scalar field
$\phi$ subject to a time dependent potential
$V(\phi)=s(t)\phi^{2}/2$, where $s(t)$ is in principle any regular function
of time (conditions on this function will be introduced later on).
The field propagates in a static background in $1+3$ dimensions whose
Cauchy surfaces are three-spheres, equipped with the standard
round metric
\begin{equation}
\label{3-sphere-metric}
h_{ab}d x^a  d x^b=d\chi^2+\sin^{2}(\chi)\left[d\theta^2
+\sin^{2}(\theta)d\sigma^2\right].
\end{equation}
Here, $\chi$ and $\theta$ have a range of $\pi$, and
$\sigma\in S^{1}$. The time coordinate $t$ runs over an
interval $\mathbb{I}$ of the real line, so that the
spacetime has the topology of ${\mathbb{I}}\times S^{3}$.
Its metric is
\begin{equation}
\label{4met}
ds^2=-dt^2+h_{ab}d x^a d x^b.
\end{equation}
In the
canonical approach, the dynamics of the system are governed by the
equations
\begin{equation}
\label{canoni-eqs}
\dot{P}_{\phi}=\sqrt{h}\left[\Delta \phi - s(t)\phi\right],
\qquad \dot{\phi}=\frac{1}{\sqrt{h}}P_{\phi},
\end{equation}
where $P_{\phi}$ is the canonical momentum of
$\phi$, $h=\sin^{2}(\theta)\sin^{4}(\chi)$ is the
determinant of the metric (\ref{3-sphere-metric}), $\Delta$
denotes the Laplace-Beltrami operator on $S^{3}$, and the dot
stands for the time derivative.

The canonical phase space of
the theory is a symplectic linear space $\Gamma$
coordinatized by the field variables $(\phi,P_{\phi})$ (evaluated on a
particular Cauchy section, e.g. the section $t=t_0$ for a given value
of time $t_0$) and endowed with a symplectic structure $\Omega$
such that these variables form a canonical pair, namely their corresponding Poisson bracket is:
\begin{equation}
\{\phi(x),P_{\phi }(y)\}=\delta^{(3)}(x-y),\end{equation}
where the Dirac delta is defined on $S^3$.
The Eqs. of motion (\ref{canoni-eqs}) amount to the linear wave equation
\begin{equation}
\label{cov-fieldequation}
\ddot{\phi}-\Delta\phi + s(t)\phi =0.
\end{equation}
Since the Laplace-Beltrami operator on $S^{3}$ is invariant under
the {rotation} group SO(4),
the above equation is clearly invariant under this group as well.
On the other hand, notice by comparison with the Klein-Gordon equation
that a nonnegative function $s(t)$ can be interpreted as an
effective nonnegative time dependent mass
$m(t)=s^{1/2}(t)$.

Owing to the field character
of the theory, the system accepts infinite nonequivalent
representations of the CCR's.
Restricting {one's} attention to representations of the Fock type, this
 freedom is encoded in the
complex structure, which is a linear symplectic map
$j:\Gamma \to \Gamma$, compatible with the symplectic structure
[in the sense that {the bilinear map}
$\Omega(j\cdot,\cdot)$ is
positive-definite], and such that $j^{2}=-1$ (see
e.g. Refs. \cite{wald,CCAP,SH}). Different choices
of complex structure select distinct, in general not unitarily
related, spaces of quantum states for the theory; thus,
physical predictions depend on the choice of $j$.

Actually, as we have mentioned, it has been proven recently that,
in the $(\phi,P_{\phi})$ description, there exists one (and
only one) subfamily of equivalent complex
structures satisfying the criteria of SO(4) invariance
and a unitary implementation of the dynamics.
Let us sketch the main steps of the proof
and explain the corresponding quantization \cite{CMV8}.
Given the invariance of the field equation under
SO(4), it is convenient to expand the field in
terms of (hyper-)spherical harmonics $Q_{nlm}$, where the
integer $n$ satisfies $n\geq0$, and the integers $\ell$ and
$m$ are constrained by $0\leq\ell\leq n$ and $|m|\leq l$
\cite{lif,GS,jantzen}. In
this basis, the Laplace-Beltrami operator $\Delta$ is
diagonal with eigenvalues equal to $-n(n+2)$.
Although the (hyper-)spherical harmonics are complex functions,
it is straightforward to obtain a real basis from the real and
imaginary parts of $Q_{nlm}$, with which one can directly expand the
real field $\phi$
(see Ref. \cite{CMV8} for details). The degrees of freedom are
represented by the coefficients $q_{n\ell m}$ in this expansion, which
can be understood as a discrete set of modes. These
are functions of time which satisfy the linear equation
\begin{equation} \label{q-eq} \ddot
q_{n\ell m}+\left[\omega_n^2+s(t)\right]q_{n\ell m}=0,
\end{equation}
with $\omega^2_n=n(n+2)$. Hence, the modes
$q_{n\ell m}$ are decoupled from each other. Besides, together with
their canonically conjugate momenta $p_{n\ell m}=\dot q_{n\ell
m}$, they form a complete set of variables in phase space. For
each fixed value of $n$, there exist $g_n=(n+1)^2$ modes with
the same dynamics, because the equation of motion is independent
of the labels $\ell$ and $m$. Obviously, the quantity $g_n$ is
just the dimension of the corresponding eigenspace of the
Laplace-Beltrami operator. The canonical phase space
$\Gamma$ can be split then as a direct sum
\begin{equation} \label{split}
\Gamma = \bigoplus_n {\cal Q}_n\oplus{\cal P}_n,
\end{equation}
where ${\cal Q}_n$ and ${\cal P}_n$ are the respective configuration
and momentum subspaces for the modes with fixed $n$. From now on, we will omit
the labels $\ell$ and $m$, unless they are necessary in the
analysis.

Furthermore, in the following we restrict our study to the inhomogeneous
sector, namely, to modes with $n\neq0$. This does not affect the
properties of the system related to its field character, because this
is maintained if one removes a finite number of
modes.\footnote{The requirements on the quantization
of the zero mode, $q_{0}$, may lead to extra conditions on the
function $s(t)$. See Ref. \cite{CMV8} and Sec. \ref{sec:conc}
for further comments on this issue.}
We will introduce the annihilation and creationlike
variables
\begin{eqnarray}
\label{basic-var} \left( \begin{array}{c} a_n \\
a_n^*\end{array}\right) = {\frac{1}{\sqrt{2\omega_n}}}\left(
\begin{array}{cc}
\omega_n & i\\
\omega_n & -i\end{array}\right)
\left(
\begin{array}{c} q_n \\ p_n\end{array}\right).
\end{eqnarray}
Notice that these are precisely the variables which one would
naturally adopt in the case of a free massless scalar field.
They form a complete set in the phase space
of the inhomogeneous sector. Given a set of initial data
$\{a_{n}(t_0),a_{n}^*(t_0)\}$ at initial time $t_0$, it is
possible to write the classical evolution to an arbitrary
time $t\in \mathbb{I}$ in the form
\begin{eqnarray}
\label{bogo-transf} \left( \begin{array}{c} a_n(t) \\
a_n^*(t)\end{array}\right) = \left( \begin{array}{cc}
\alpha_n(t,t_0) &\beta_n(t,t_0)\\
\beta_n^*(t,t_0) & \alpha_n^*(t,t_0)\end{array}\right)
\left(
\begin{array}{c} a_n(t_0) \\ a_n^*(t_0)\end{array}\right).
\end{eqnarray}
Let us call ${\cal U}_n(t,t_0)$ the linear evolution operator
defined in this way.
Since the time functions $\alpha_n(t,t_0)$ and
$\beta_n(t,t_0)$ provide a symplectomorphism on $\Gamma$,
one has
\begin{equation}
|\alpha_n(t,t_0)|^2-|\beta_n(t,t_0)|^2=1,
\end{equation}
independently of the particular values of $n$, $t_0$, and
$t$. Such Bogoliubov coefficients $\alpha_n(t,t_0)$ and $\beta_n(t,t_0)$ of
this evolution map can be determined in the way explained
in Ref. \cite{CMV8}. For our present analysis, we only
need to employ that their asymptotic behavior when $n\to \infty$ is
given by
\begin{equation}
\label{abeta}
\alpha_n(t,t_0)= e^{-i(n+1)\tau}+{O\left(\frac{1}{n}\right)},
\qquad \beta_n(t,t_0)= {O\left(\frac{1}{n^2}\right)},
\end{equation}
where $\tau=t-t_0$ and the symbol
{$O$ denotes the asymptotic order}.
The derivation of this asymptotic behavior makes use of the mild
assumption that the function $s(t)$ in Eq. (\ref{cov-fieldequation})
must be differentiable, with a derivative that is
integrable in every closed subinterval of $\mathbb{I}$.
To simplify the notation, we will
omit in the following the reference to the initial time $t_0$
in the coefficients of the evolution operator and in the
initial data, called now $\{a_{n},a_{n}^*\}$.

The SO(4) symmetry of the field equations is imposed at the
quantum level by demanding that the complex structure be invariant
under this group. We call {\sl invariant} this class of complex structures.
By Schur's lemma \cite{schur}, any invariant complex structure has to be
block diagonal with respect to the decomposition of the phase space
as the direct sum of the subspaces ${\cal Q}_n \oplus{\cal P}_n$.
In other words, the complex structure can be  decomposed as a direct sum
$j=\bigoplus_n j_n$, where $j_n$ is an invariant complex structure
defined on the $n$-th subspace of the inhomogeneous sector of
$\Gamma$. Moreover, a further application of Schur's lemma shows that
each of the complex structures $j_n$ is
again block diagonal and independent of the labels $l$
and $m$, so that it can be characterized by a complex structure
in two dimensions, describing e.g. the action on the annihilation and
creationlike variables $(a_{n}, a^*_{n})$ for any fixed mode
labels $l$ and $m$ (see Ref. \cite{CMV8} for more details).

On the other hand, let us call $j_0$
the complex structure which in our basis of variables
$\{a_{n}, a^*_{n}\}$ takes the diagonal form:
\begin{eqnarray}
{j_0}_n= \left( \begin{array}{cc} i & 0
\\ 0 & -i \end{array} \right).
\end{eqnarray}
A general
complex structure $j$ is related with $j_0$
via a symplectic transformation
$j={\cal K} j_0 {\cal K}^{-1}$. Taking into account the
form of the invariant complex structures,
the symplectic transformation ${\cal K}$ must be also block diagonal
and independent of the degeneracy labels $l$ and $m$.
We call ${\cal K}_n$ the $2\times 2$ block corresponding to the $n$-th
mode, for which we adopt the notation:
\begin{eqnarray}
\label{symplect} {\cal K}_n = \left( \begin{array}{cc}
\kappa_n &\lambda_n\\
\lambda_n^* & \kappa_n^*\end{array}\right).
\end{eqnarray}
The symbol $*$ denotes again complex conjugation.
Here, $|\kappa_n|^2-|\lambda_n|^2=1$ because ${\cal K}_n$ is a
symplectomorphism.
It follows in particular that
$|\kappa_n|\geq 1$ $\forall n\in\mathbb{N}^+$. Note
that there exist infinite invariant complex
structures. Actually, they are not all unitarily equivalent, so that
the imposition of SO(4) symmetry does not
eliminate the ambiguity in the Fock quantization on its own.

In order to select a class of equivalent invariant complex
structures we need to appeal to additional conditions. A
unitary implementation of the classical dynamics at the
quantum level turns out to determine a preferred class,
and hence a unique Fock quantization up to equivalence.
We recall that a symplectic transformation $T$ is
implementable as a unitary transformation in the quantum
theory for a given complex structure $j$ if and only if $j -
TjT^{-1}$ is a Hilbert-Schmidt operator
(on the one-particle Hilbert space defined by $j$, see e.g. Refs.
\cite{a-m,SH}). In the case of the time evolution operator,
and choosing the complex structure $j_0$,
this condition is satisfied if and only if
\begin{equation}
\label{SQS}
\sum_{n\ell
m}|\beta_{n}(t)|^2=\sum_ng_n|\beta_{n}(t)|^2<\infty
\quad \forall t\in\mathbb{I},
\end{equation}
i.e., if and only if the sequences $\{\sqrt{g_n}\beta_{n}(t)\}$ are square
summable (SQS)
for all possible values of time.
Then, since $\sqrt{g_n}=n+1$,
the asymptotic behavior (\ref{abeta}) of the beta coefficients
guarantees the desired summability, ensuring that the dynamics is
implemented unitarily in the Fock quantization picked up by $j_0$,
namely the complex structure associated to the natural choice of
annihilation and creationlike variables for the free massless case.

Let us suppose now that we choose a different
invariant complex structure $j$, which can be obtained from $j_0$ by
means of a symplectic transformation ${\cal K}$, as we have commented.
The unitary implementation of the evolution operator with respect to
the new complex structure $j$ is equivalent to the unitary implementation
of a transformed evolution operator with respect to the complex structure
$j_0$ \cite{CMV8}. This transformed evolution
operator is obtained from the original one by the action of ${\cal K}$.
Its diagonal blocks are ${\cal K}_n{\cal U}_n(t){\cal K}_n^{-1}$, with
corresponding beta coefficients given by
\begin{equation}
\label{bjn}
\beta^j_n(t):=(\kappa_n^*)^2\beta_n(t)-\lambda_n
^2\beta_n^*(t)+2i\kappa_n^*\lambda_n
\Im[\alpha_n(t)].
\end{equation}
Here, the symbol $\Im$ denotes imaginary part.
If one assumes that the evolution is unitary in the Fock quantization
determined by $j$, so that the sequences
$\{\sqrt{g_n}\beta^j_{n}(t)\}$ are SQS
$\forall t\in \mathbb{I}$, one can prove that the sequence
$\{\sqrt{g_n}\lambda_n\}$ must be SQS as well
\cite{CMV8}. But this summability is precisely the
sufficient and necessary condition for the unitary implementation
of the symplectic transformation ${\cal K}$ in the Fock
representation determined by $j_0$, what amounts to the
equivalence of the two complex structures $j$ and $j_0$.
Therefore, the SO(4) invariance and the requirement of unitary
dynamics select a unique equivalence class of complex structures,
removing the ambiguity in the choice of representation of the CCR's.

\section{The ambiguity in the choice of field and the unitarity criterion}
\label{sec:trans}

Although we have succeeded in selecting a preferred representation of
the CCR's for the $(\phi,P_{\phi})$ variables, we can always change
from the $(\phi,P_{\phi})$ description of the phase space to a new
canonical description by means of a canonical transformation. Since many
canonical transformations fail to be represented by unitary operators
quantum mechanically, the classical equivalence of these descriptions
may be broken in the quantum arena, originating another type of
ambiguity in the quantization. In our case, we are only interested
in considering linear, local canonical transformations, which respect
the linearity of the field equations and, consequently, the linear
nature of the structures of the system. As we have explained in the
Introduction, the class of canonical transformations that we want to
analyze results in a time dependent scaling of the field.
It is this time dependence what makes the transformation nontrivial;
otherwise, any admissible representation of the original field would
provide an admissible one for the transformed field by linearity.
But when the canonical transformation is time dependent, the field
dynamics changes, affecting the properties of the quantum theory.

A time dependent scaling of the configuration field
variable can be regarded as a contact transformation, which can
easily be completed into a canonical one. Then, the canonical
momentum must experience the inverse scaling and, optionally, may
be modified with the addition of a term depending on the
configuration field variable, which
we restrict to be linear (and local), according to our previous
comments. The coefficient in this linear term
may vary in time, like the rest of coefficients in the
linear canonical transformation under consideration.
In this way, one obtains a transformation of the form
\begin{equation}
\label{time-dep-ct}
\varphi=F(t)\phi,\qquad
P_{\varphi}=\frac{P_{\phi}}{F(t)}+G(t)\sqrt{h}\phi.
\end{equation}
We recall that the momentum variable is a scalar density of
unit weight. This explains the square root of the determinant
of the spatial metric appearing in Eq. \eqref{time-dep-ct}.
In order that the transformation does not spoil the differential
formulation of the field theory, nor produces singularities,
$F$ and $G$ are restricted to be two real and differentiable
functions of time, with $F(t)$ different from zero everywhere.
Notice also that the {homogeneity} of $F$ and $G$ preserves
the SO(4) invariance of the field dynamics. In the following,
we will consider only time dependent canonical transformations
of the form (\ref{time-dep-ct}).

As we have pointed out, different choices of the basic fieldlike
variables typically lead to distinct dynamics.
For instance, a canonical transformation with
{$F(t)=1/a(t)$}, where $a$
is a solution of the second order differential equation
\begin{equation}
\frac{\ddot{a}}{a}-m^{2}a^{2}+s(t)=0,
\end{equation}
leads from
the field equation (\ref{cov-fieldequation}) to the dynamics
of a Klein-Gordon field with mass $m$ propagating in
the FRW background
$d\tilde{s}^2=a^{2}(t)ds^2$ [see Eq. (\ref{4met})].
This indicates that one can extract information about the
dynamics of different field theories by performing a time dependent
canonical transformation of the above type. Let us emphasize,
however, that with this procedure one is not transforming a
given field theory into another one, but rather considering
distinct field descriptions of a given physical system,
assuming that none of these descriptions is imposed from the start.
In this kind of systems, one has to address the ambiguity associated
with the choice of field parametrization (i.e., with the selection
of {\sl fundamental} field, together with its associated dynamics).
Then it is necessary to invoke
additional, physically acceptable criteria to pick up
a preferred quantization; otherwise, the
significance of the predictions of the quantum theory would be
in question. The criteria that we are going to adopt are indeed
the same that allow us to select a unique
equivalence class of Fock representations in the
$(\phi,P_{\phi})$ representation, that is,
the SO(4) invariance and the unitary implementation of
the evolution.

For the rest of our analysis, it is convenient to split the time
dependent canonical transformation \eqref{time-dep-ct}
into two parts, one that takes care of the initial conditions on the
functions $F(t)$ and $G(t)$, and the other that carries all the time
dependence. We fix once and for all an initial reference time $t_0$,
and denote by $F_0$ and $G_0$, respectively, the initial values
$F(t_0)$ and $G(t_0)$. Then, any transformation of the form
\eqref{time-dep-ct} can be obtained
as the composition of the canonical transformation
\begin{equation}
\label{const-ct}
\tilde\varphi=F_0\varphi,\qquad
P_{\tilde\varphi}=\frac{P_{\varphi}}{F_0}+G_0\sqrt{h}\varphi,
\end{equation}
which does not vary in time,
with a linear canonical transformation of the type \eqref{time-dep-ct},
\begin{equation}
\label{time-vary-ct}
\varphi=f(t)\phi,\qquad
P_{\varphi}=\frac{P_{\phi}}{f(t)}+g(t)\sqrt{h}\phi,
\end{equation}
but such that the functions $f(t)$ and $g(t)$ now have
fixed initial values, namely
$f(t_0)=1$ and $g(t_0)=0$.
The transformation \eqref{const-ct}
is just a time independent linear one, with no impact
in our discussion, given the linearity of the Fock
representations of the CCR's.
In fact, if a quantization with SO(4) invariance and unitary dynamics
is achieved for the canonical pair $(\varphi,P_{\varphi})$,
one immediately obtains a quantization with the same properties
for the transformed pair
$(\tilde\varphi,P_{\tilde\varphi})$.
No real
ambiguity comes from this kind of transformations, since the
quantum representation for the original and the transformed
fields is actually the same
(see Ref. \cite{CMV5} for more details).

Thus, we shall  restrict our analysis to the family of
canonical transformations
\eqref{time-vary-ct} with fixed initial conditions.
We will demonstrate that any such transformation,
except the identity, leads to a classical evolution which admits
no unitary implementation with respect to any of the
Fock representations defined by an SO(4) invariant complex
structure. Thus, our criteria fix completely the choice of field
description
[up to a trivial time independent transformation of the type
\eqref{const-ct}].

Let us discuss now the form of the new dynamics
obtained with the transformation \eqref{time-vary-ct}, and present
the mathematical condition necessary for a unitary implementation
of this dynamical evolution. We recall that the
linear transformation \eqref{time-vary-ct} preserves the SO(4)
invariance of the field equations and that we demand that the (real)
functions $f(t)$ and $g(t)$ be differentiable.
Moreover, [like the functions $F(t)$] the function $f(t)$
is required to differ from zero everywhere. The sign of the function
$f(t)$ is therefore constant and, since its initial value has been fixed
equal to the unit, in what follows we take
$f(t)>0$ $\forall t\in \mathbb{I}$.

As we have already commented, since the canonical transformation
\eqref{time-vary-ct} depends on time, the classical evolution operator
that describes the dynamics of the  pair $(\varphi,P_{\varphi})$
differs from that corresponding to the original pair $(\phi,P_{\phi})$.
In order to describe the new dynamics, we will follow the same
procedure adopted in the previous section. Namely, we first expand the
field $\varphi$ and its momentum $P_{\varphi}$ in (hyper-)spherical
harmonics, extracting in this way their spatial dependence,
and then introduce annihilation and creationlike variables, defined
in terms of the coefficients of the expansion like in
Eq. (\ref{basic-var}). One can check [using the transformation
\eqref{time-vary-ct} and the corresponding initial conditions]
that, with those variables, the blocks of the original
evolution matrix ${\cal U}_n(t)$ are replaced by new $2\times 2$
matrices $\tilde {\cal U}_n(t) = {\cal T}_n(t){\cal U}_n(t)$,
where\footnote{While the dependence of ${\cal U}_n(t)$
on $t_0$ is not shown explicitly to simplify the notation,
the matrix ${\cal T}_n(t)$ actually does not depend on the
initial time.}
\begin{eqnarray}
\label{evol-trans}
{\cal T}_n(t) &:=&
\left(\begin{array}{cc} f_{+}(t)+i\frac{g(t)}{2\omega_n} \;&
f_{-}(t)+i\frac{g(t)}{2\omega_n} \\ f_{-}(t)-
i\frac{g(t)}{2\omega_n}\;& f_{+}(t)-i\frac{g(t)}{2\omega_n}
\end{array} \right)
\end{eqnarray}
and $2 f_{\pm}(t):=f(t)\pm 1/f(t)$.
Finally, a straightforward computation allows us to obtain the
Bogoliubov coefficients $\tilde \alpha_n(t) $ and $\tilde \beta_n(t)$
of the evolution matrices  $\tilde
{\cal U}_n(t)$, which are of the form
\begin{eqnarray}
\label{Bog-relat}
\tilde \alpha_n(t)   &:=& f_+(t)\alpha_n(t) +
f_-(t)\beta^*_n(t) + i
\frac{g(t)}{2\omega_n}[\alpha_n(t)+\beta_n^*(t)],
\nonumber\\
\tilde \beta_n(t)   &:=& \! f_+(t) \beta_n(t) +
f_-(t)\alpha^*_n(t)+i\frac{g(t)}{2\omega_n}[
\alpha^*_n(t)+\beta_n(t)].
\end{eqnarray}

One can now simply follow the procedure explained in
Sec. \ref{sec:static} and write down the condition for a
unitary implementation
of the dynamics of the transformed canonical pair
$(\varphi,P_{\varphi})$ with respect to a representation of the CCR's
defined by an SO(4) invariant complex structure. We again call
${\cal K}$ the symplectic transformation that determines the invariant
complex structure $j$ under consideration in terms of the complex
structure of reference $j_0$. We also adopt the notation
\eqref{symplect} for its coefficients, which do not depend on time.
The new dynamics admits a unitary implementation with respect
to the representation determined by $j$ if and only if the sequences
$\{\sqrt{g_n}\tilde\beta^J_n(t)\}$ are SQS $\forall t \in \mathbb{I}$,
where
\begin{equation}\label{tilde:beta}
\tilde\beta^j_n(t):=(\kappa_n^*)^2\tilde\beta_n(t)-\lambda_n
^2\tilde\beta_n^*(t)+2i\kappa_n^*\lambda_n
\Im[\tilde\alpha_n(t)].
\end{equation}

\section{Uniqueness in the choice of field description}
\label{sec:uniq}

We will now present the detailed proof that the unitarity condition
introduced in the
previous section implies that the transformation
\eqref{time-vary-ct} must in fact be the identity transformation.

Let us assume that the unitarity condition is satisfied. Then,
the sequences $\{\sqrt{g_n}\tilde\beta^j_n(t)\}$ are SQS
for all values of time in the considered interval $\mathbb{I}$.
In particular, this requires that the terms
$\sqrt{g_n}\tilde\beta^j_n(t)$ of these sequences tend to
zero in the limit $n\to\infty$. Since both $g_n$
and $|\kappa_n|$ are greater than 1, it must be also true that
$\tilde\beta^j_n(t)/(\kappa_n^*)^2$ tends to zero.
Taking into account the asymptotic
limits of the Bogoliubov coefficients $\alpha_n(t)$ and $\beta_n(t)$,
given in Eq. \eqref{abeta}, and introducing for convenience the notation
$z_n=\lambda_n/\kappa_n^*$, we arrive at the conclusion that
\begin{equation}
\label{lead-seq}
\left[e^{i(n+1)\tau}-z_n^2e^{-i(n+1)\tau}\right]
f_-(t)
-2iz_n\sin[(n+1)\tau]f_+(t)
\end{equation}
must have a vanishing limit when $n\to\infty$ for all values of $t\in\mathbb{I}$.
Recall that $\tau=t-t_0$.

By considering separately the real and imaginary parts of the above
expression, we get that the two sequences given respectively by
\begin{eqnarray}
\label{real-lead-seq}&& \left(2\Im\left[z_n\right] f_+(t) -
\Im\left[z_n^2\right]
f_-(t)\right)\sin[(n+1)\tau]\nonumber\\
&&+\left(1-
\Re\left[z_n^2
\right]\right)f_-(t)\cos[(n+1)\tau]
\end{eqnarray}
and
\begin{eqnarray}
\label{im-lead-seq} &&\left(\left\{1+ \Re
\left[z_n^2\right]
\right\}f_-(t)- 2\Re\left[z_n\right]f_+(t)\right)\sin[(n+1)\tau]
\nonumber\\
&&-\Im
\left[z_n^2\right]
f_-(t)\cos[(n+1)\tau]
\end{eqnarray}
have to tend to zero when $n\to \infty$ $\forall
t\in\mathbb{I}$. Here, the symbol $\Re$ denotes real part.

We can now apply arguments similar to those presented in
Ref. \cite{CMV5} and
show that, if it is true that the sequences given in
Eq. (\ref{im-lead-seq}) tend to zero for all
values of time, then it is impossible that the two sequences
formed by
\begin{equation}
\label{coeff}
1-\Re\left[z_n^2\right]\quad \text{and}\quad\Im\left[z_n^2\right]
\end{equation}
have simultaneously a vanishing limit on any (infinite)
subsequence of the positive integers $\mathbb{M}\subset\mathbb{N}^+$
(i.e., for $n\in\mathbb{M} \subset\mathbb{N}^+$). Let us see this
in more detail.

We first note that $\left({\Re}[z_n]\right)^2$
tends to the unit whenever the two terms in Eq. (\ref{coeff})
tend to zero. This can be checked by summing the
square of the two terms (\ref{coeff}),
which gives
\begin{equation}
(1-|z_n|^2)^2+4\left({\Im}[z_n]\right)^2.
\end{equation} By our assumptions,
this expression tends to zero on a given subsequence
$\mathbb{M}$. Then, we get that $|z_n|$ must
tend to the unit and ${\Im}[z_n]$ to
zero on this subsequence, what implies that
the limit of $\left({\Re}[z_n]\right)^2$ is equal to one.

Suppose then that there really exists
a particular subsequence $\mathbb{M}\subset \mathbb{N}^+$ such
that the terms (\ref{coeff}) tend to
zero on it for all possible values of time.
Since the factor
\begin{equation}
f_-(t) \cos[(n+1)\tau],
\end{equation}
which multiplies ${\Im}[z_n^2]$
in Eq. (\ref{im-lead-seq}), is bounded for
every {particular} value of $t$, it follows that
\begin{equation}
\left(\left\{1+ \Re
\left[z_n^2\right]
\right\}f_-(t)- 2\Re\left[z_n\right]f_+(t)\right)
\sin[(n+1)\tau]
\end{equation}
must have a vanishing limit on $\mathbb{M}$ $\forall t\in\mathbb{I}$.
Besides, since by hypothesis
$1-{\Re}\left[z_n^2\right]$
tends to zero on $\mathbb{M}$ as well, one further obtains that
\begin{equation}
\label{n13} \left(f_-(t) -{\Re}\left[z_n\right]f_+(t)
\right)\sin[(n+1)\tau]
\end{equation}
must tend to zero on $\mathbb{M}$ at each possible value of the time
$t$.

We now make use of the result proven above that
$\left({\Re}[z_n]\right)^2$ necessarily tends
to the unit on $\mathbb{M}$. Then, there exists at least one
subsequence $\mathbb{M}'\subset \mathbb{M}$ such that
${\Re}[z_n]$ tends to $1$ or to $-1$ on
$\mathbb{M}'$. In any of these cases, given that $\mathbb{M}'$
is a subsequence of $\mathbb{M}$, and hence
expression (\ref{n13}) must tend to zero on $\mathbb{M}'$,
we conclude (using the definition of $f_{\pm}$) that either
\begin{equation}
\sin[(n+1)\tau]f(t)\qquad {\rm or}
\qquad \frac{\sin[(n+1)\tau]}{f(t)}\end{equation}
(or both) have a vanishing limit on the subsequence $\mathbb{M}'\subset
\mathbb{N}^+$ $\forall t\in\mathbb{I}$. But, since the function $f(t)$
is continuous and vanishes nowhere, this implies that $\sin[(n+1)\tau]$
must tend to zero on $\mathbb{M}'$ for all possible values of time in
$\mathbb{I}$, or equivalently $\forall \tau\in\mathbb{\bar I}$, where
$\mathbb{\bar I}$ is the domain obtained from $\mathbb{I}$ after a shift
by the initial time $t_0$.

Let us finally prove that this limiting behavior is not allowed.
Take a positive number $L$ such that
$[0,L]\subset \mathbb{\bar I}$. We have, in particular,
that $\sin^2[(n+1)\tau]$ tends to zero on $\mathbb{M}'$
$\forall \tau\in[0,L]$. However, a simple application of
the Lebesgue dominated convergence shows that this statement is
false. The details are presented in the Appendix. Essentially,
one can see that the integral of $\sin^2[(n+1)\tau]$ over the interval
$[0,L]$ is bounded from below by a strictly positive number for
large $n$, something which is incompatible with a vanishing limit
for this function in the entire interval.
Therefore, one can exclude the possibility that the two
sequences of time independent terms appearing
in  Eq. (\ref{coeff}) can both converge to zero on
a subsequence $\mathbb{M}'\subset \mathbb{N}^+$.

We will now use this fact to demonstrate that the
function $f(t)$ is necessarily a constant function.
Let us study again the real sequences given by
Eqs. (\ref{real-lead-seq}) and
(\ref{im-lead-seq}) which,
as we have seen, must necessarily
tend to zero in the limit $n\to\infty$ for
all possible values of time $t\in\mathbb{I}$ if the
dynamics of the $(\varphi,P_{\varphi})$
canonical pair admits a unitary implementation.

We concentrate our attention on a specific subset of values of
the shifted time $\tau$, namely, all values of the
form $\tau=2\pi q/p$ where $q$ and $p$ can be any
positive integers, except for the
condition that the resulting value of $\tau$ belongs to
the interval of definition of this variable, $\mathbb{\bar I}$.
For each value of $p$, we consider the subsequence of positive
integers
\begin{equation}
\mathbb{M}_p:=\{n=kp-1>0,\ k\in \mathbb{N}^+\}.
\end{equation}
Given $p$, the terms (\ref{real-lead-seq}) and (\ref{im-lead-seq})
tend to zero on the subsequence $\mathbb{M}_p$
when $n\to \infty$ for all the values
of $\tau$ reached when $q$ varies. Then, we reach the
conclusion that both
\begin{equation}\label{lead-seq-simpl}
\left(1-\Re\left[z_{kp-1}^2\right] \right)f_-\left(t_0+\frac{2\pi
q}{p}\right)
\end{equation}
and
\begin{equation}
\label{lead-seq-simpl2}
\Im
\left[z_{kp-1}^2\right]f_-\left(t_0+\frac{2\pi q}{p}\right)
\end{equation}
must tend to zero as $k$ goes to infinity. The limit must vanish for
every possible integer value of $p$ and $q$. Note however that the
time independent factors on the left of these expressions are precisely
those given in Eq. (\ref{coeff}), which
we have proven that cannot tend simultaneously to zero on any
subsequence of the positive integers, e.g. those provided by
$\mathbb{M}_p$ for each of the values of $p$. Therefore, the only
possibility left is that the function $f_{-}(t_0+2\pi q/p)$ is
equal to zero at all the considered values of $p$ and $q$.
Using the fact that $f(t)>0$ $\forall t\in\mathbb{I}$,
the last result amounts to the equality
\begin{equation}
f\left(t_0+\frac{2\pi q}{p}\right)=1\qquad \forall p,q.
\end{equation}
Realizing that the subset of time values
$\{t_0+2\pi q/p\}$ is dense in $\mathbb{I}\subset\mathbb{R}$
and that the function $f(t)$ is continuous,
we are led to the conclusion that $f(t)$ must equal
the unit function on its entire domain.

It remains to be proven that the function $g(t)$ in
the transformation \eqref{time-vary-ct} necessarily vanishes,
under the condition of unitary dynamics. Note first that the
identity $f(t)=1$ that we have just demonstrated implies that
$z_n$ tends to zero when $n\to\infty$. In fact, after introducing
this identity in Eq. (\ref{lead-seq}), one sees that the
sequences $\{z_n\sin[(n+1)\tau]\}$ must tend to zero
$\forall \tau\in\mathbb{\bar I}$.
Therefore, in order to avoid again the false conclusion that
$\sin^2[(n+1)\tau]$ tends to zero on some subsequence of the
positive integers for all values of $\tau$ in a
compact interval, it is necessary that the complex sequence
$z_n$ has a vanishing limit. Taking into account that
$|\kappa_n|^2=|\lambda_n|^2+1$, it is straightforward to
check that the sequence formed by the coefficients $\lambda_n$
must tend to zero, and that $1/|\kappa_n|^2$ (and $|\kappa_n|^2$)
approaches the unit in the limit of large $n$, what implies
in particular that the sequence given by $\kappa_n$ is bounded.

To complete the proof that $g(t)$ vanishes,
we consider again the sequences
$\{\sqrt{g_n}\tilde\beta^j_n(t)\}$, particularized now to the only
value allowed for the function $f(t)$, namely the identity,
so that $f_{+}(t)=1$ and $f_{-}(t)=0$.
Employing the definition of the coefficient $\beta^j_n(t)$, given in Eq.
(\ref{bjn}), one can check that the leading terms in $\tilde\beta^j_n(t)$ are
\begin{equation}
\tilde\beta^j_n(t) \cong
\beta^j_n(t)+i\frac{g(t)}{2\omega_n}
\left[(\kappa_n^*)^2\alpha^*_n(t)+\lambda_n^2\alpha_n(t)\right].
\end{equation}
The condition of unitarity demands that $\sqrt{g_n}\tilde\beta^j_n(t)$ tend
to zero in the limit  $n\to\infty$ at all values of time,
and therefore the same must happen to the sequences with terms
$\sqrt{g_n}\tilde\beta^j_n(t)/{\kappa^*_n}^2$.
Using this condition and taking into account the known
asymptotic behavior \eqref{abeta} of $\alpha_n(t)$ and $\beta_n(t)$, as well that
$\lambda_n$ tends to zero and $\sqrt{g_n}/{\omega_n}$ tends to the unit
for large $n$, a simple calculation leads
to the result that the sequences
given by
\begin{equation}
\label{z2} g(t)-4
z_n
\omega_n\sin[(n+1)\tau]e^{-i(n+1)\tau}
\end{equation}
must have a vanishing limit $\forall t\in\mathbb{I}$.
We then consider the real and imaginary parts of these sequences,
namely
\begin{equation}\label{z2_simpl}
g(t)-4|z_n|\omega_n\sin[(n+1)\tau]\cos[(n+1)\tau-\delta_n]
\end{equation}
and
\begin{equation}\label{z2_simpl2}
4|z_n|\omega_n\sin[(n+1)\tau]\sin[(n+1)\tau-\delta_n],
\end{equation}
where we have
written the complex numbers $z_n$  in terms of its
phase and complex norm:
\begin{equation}
z_n=|z_n| e^{i\delta_n}.
\end{equation}
Although we already know that $|z_n|$ tends to zero,
the limit of the product $|z_n|\omega_n$ is still undetermined,
because $\omega_n$ grows like $n$ at infinity.

Let us suppose first that the sequence  $\{|z_n|\omega_n\}$ tends to zero.
In this case, recalling that the sequences given in Eq. (\ref{z2_simpl})
should
tend to zero $\forall t \in \mathbb{I}$,
it follows immediately that $g(t)$ must be the zero function on $\mathbb{I}$,
as we wanted to prove. Finally, let us demonstrate that the alternate
possibility, i.e. the hypothesis that $\{|z_n|\omega_n\}$ does not tend to zero,
leads to a contradiction. We make use of the fact that the sequences
formed by the terms (\ref{z2_simpl2}) tend to zero $\forall t$.
If $\{|z_n|\omega_n\}$ does not tend to zero, there must exists
a subsequence $\mathbb{M}$ of the positive integers such that the
(positive) sequence $\{|z_n|\omega_n\}$ is bounded from below on $\mathbb{M}$.
Thus, on that subsequence,
\begin{equation}
\sin[(n+1)\tau]\sin[(n+1)\tau-\delta_n]
\end{equation}
must necessarily have a zero limit $\forall \tau\in \mathbb{I}$.
But, as shown also in the Appendix, this last statement can never be true.
Again the crucial argument involves the application of the Lebesgue
dominated convergence.

As a result, the only function $g(t)$ that is allowed by the condition
of unitarity is the zero function. In total, we have demonstrated that
the only canonical transformation of the type (\ref{time-vary-ct})
which is permitted once one accepts the unitarity criterion is the
trivial one, i.e. the identity transformation. In this way, the choice
of a field parametrization for the system turns out to be {\sl completely}
fixed (up to irrelevant constant scalings) by the requirements of
invariance under the symmetry group of the field equations, SO(4), and
the unitary implementation of the dynamics. The ambiguity in the selection
of a field description is totally removed.

\section{Conclusions and discussion}
\label{sec:conc}

In this work , we have begun our analysis
by reviewing the Fock quantization of a
scalar field with a time varying mass in a static background, in which
the inertial spatial sections have $S^3$ topology. For this particular
scenario,  we have seen that the criteria of: \/{\it i}) invariance of the
vacuum under the SO(4) symmetry
of the field equations; and \/{\it ii}) unitary implementation of the
field dynamics,
are sufficient  to select a unique Fock representation of the CCR's.

An additional question concerns the possibility of
changing the field
description, if one allows for a scaling of the field by time dependent
functions. This is a situation frequently found in cosmology, where it
is common to introduce scalings of the fields in order to absorb
part of the time dependence of the cosmological background. The
prototypical example is that of fields in an FRW spacetime with compact
topology ($S^3$ for our discussion), or the closely related scenario of
field perturbations around an FRW background of that kind.
There is therefore an extra  ambiguity affecting the quantization of
such systems, namely the choice of the field description, which necessarily
affects the dynamics.

In the above mentioned systems,
it is generally the case that a time dependent scaling of the field
renders the field equations into a form describing the effective
propagation in a static background with a time varying mass, i.e.
the model that we considered initially. We have
demonstrated here the result that we anticipated in Ref. \cite{CMOV},
namely, that our criteria of symmetry invariance and unitary
evolution allow only for one admissible field description among all
those that can be reached by means of time dependent canonical
transformations that include a time dependent scaling of the field.
The analyzed canonical transformations are linear, in order to maintain
the linearity of all the structures on phase space, and preserve the
symmetry of the field equations.

To arrive at this uniqueness result, very mild requirements
have been imposed on the mass function $s(t)$ appearing in the
field equation \eqref{cov-fieldequation}. Specifically, the
only condition that has been assumed is that the mass function has
a first derivative which is integrable in all closed subintervals of
the domain of definition. In addition, if one wants that the zero mode
of the scalar field (the homogeneous sector) can be quantized
consistently in the standard Schr\"odinger representation with
the Lebesgue measure (on $\mathbb{R}$), an extra condition has to
be added: the mass $s(t)$ has to be \textit{nonnegative} for all
possible values of time.

Let us comment on some key points underlying
our uniqueness
result. A fundamental question is to understand why one can reach
unitarity in the
quantum evolution and how this unitarity selects a
unique field
description as well as a unique equivalence class of complex
structures for it, among the set of all symmetry invariant
complex structures. In this respect, we first notice that infrared
divergences are not an issue to begin with, owing to the fact
that the spatial sections have compact topology (leading in particular  to a discrete
spectrum for the Laplace-Beltrami operator). Like for
many other considerations in cosmology, the compactness of the
spatial sections is essential. When the spatial topology is not
compact, the infrared problem appears and changes
the scenario drastically.\footnote{For instance, the well known inequivalence
of the quantum representations corresponding to free scalar fields of different masses
in Minkowski spacetime is precisely due to the long range behavior of the quantum fields.
See Ref. \cite{MTV} for an account.} On the other hand, the ultraviolet
divergences are
absent in the system precisely because we are using an appropriate representation
of the CCR's. This representation turns out to be the one
naturally associated with  a free massless scalar field. The reason is that,
in the asymptotic limit of large wavenumbers, which are the
 relevant modes   for the ultraviolet regime,
the behavior of the system (when the field is properly scaled)
approaches sufficiently fast the behavior of a massless field.
Only Fock quantizations (with the desired invariance)
which are equivalent to the one that we have chosen keep this good
ultraviolet property. In this way, one
obtains a single family of unitarily equivalent Fock quantizations
which incorporate
the symmetries of the field
equation and respect the unitarity in the evolution.
Concerning the choice of field description, let us also note
that nonstationary spacetimes give rise to damping terms
(first order time derivatives of the scalar field) in the
equations of motion. In relation with our previous
comments, such contributions spoil the unitary
implementation of the dynamics at the quantum level.
Fortunately, a suitable scaling of the field relegates
all the information about the
nonstationarity of the system to the
(effective) mass term.

Let us see this last point in some more detail. As we have explained,
in order to quantize a scalar field in a nonstationary setting
along the lines presented in this paper, one generically performs
a canonical transformation which involves a time dependent
scaling of the field, so that the transformed field effectively
propagates in a static background.
Let us call $\varphi$ and ${\tilde s}(t)$, respectively,
the field and its time dependent mass previous to the discussed
transformation. As commented above, the corresponding field equation
contains a damping term, which is linear in $\dot{\varphi}$.
We call $r(t)$ the function multiplying $\dot{\varphi}$ in this damping
contribution. We now want to give the explicit expressions of the
time dependent scaling factor, $F(t)$ [see Eq \eqref{time-dep-ct}],
and of the mass function $s(t)$ for the field $\phi=\varphi/F(t)$.
A straightforward calculation shows that
\begin{equation}
F(t)=F_0\exp\left[-\int_{t_0}^t d\tau \frac{r(\tau)}{2}\right], \quad
s(t)={\tilde s}(t)-\frac{[r^2(t)+2\dot r(t)]}{4}.
\end{equation}
We also note that the condition imposed on $s(t)$ for the validity
of our uniqueness result is met, for instance, if ${\tilde s}(t)$
satisfies the same condition and $r(t)$ has a second derivative
which is integrable in all compact subintervals of the time domain
$\mathbb{I}$.
On the other hand, the positivity of the mass function (for a
standard quantization of the homogeneous sector) amounts just to
\begin{equation}{\tilde
s}(t)\geq\frac{[r^2(t)+2\dot r(t)]}{4}\qquad \forall t\in
\mathbb{I}.
\end{equation}

Let us now address possible generalizations of our results,
starting with the case of scalar fields in different compact
spatial manifolds. The analysis carried out here, together with the
dimensional arguments
explained in Ref. \cite{CMV8} in relation to the uniqueness of the
representation of the CCR's for the field description selected
by our criteria, strongly indicate that the results that we have
achieved for the three-sphere can be extended to other
compact spatial manifolds provided that the spatial dimension $d$
is equal or smaller than three. Suppose that, in these cases,
the representation of the
symmetry group of the field equation is irreducible in each of the
eigenspaces of the Laplace-Beltrami operator [like it happens for
SO(4) in the case of the three-sphere]. This property
is actually {\sl sufficient} (though not necessary) to
characterize the invariant complex structures in a block diagonal
form similar to that discussed in this work. One can then
follow the same kind of steps that have allowed us to complete the
proof of uniqueness, reaching analogous conclusions.

In all the cases with $d\leq3$,
our arguments therefore support
the expectation that, when one adopts
the scalar field description with propagation
in a static background, the free massless representation
provides the unique (equivalence class of)
Fock quantization that satisfies our criteria of
symmetry invariance and unitary dynamics \cite{CMV8}.
Besides, our criteria are expected to
fix again the function $f(t)$ in the  canonical
transformations of the type (\ref{time-vary-ct}). This
ensures that there is no ambiguity in the scaling of the field, either.
The only freedom remaining in the canonical transformation is
given by the function $g(t)$. It is not difficult to realize,
repeating the arguments discussed here,
that whether or not the function $g(t)$ is fixed to vanish
depends on the square summability of the sequence
$\{\sqrt{g_n}/\omega_n\}$. If the sequence is not SQS, as it happens
for the cases of the two-sphere and the three-sphere,
the function $g(t)$ must vanish. However, if the
sequence is SQS, there exists an arbitrariness and our criteria
do not determine the definition of the momentum $P_\varphi$ completely.
For instance, this is the case of the circle $S^1$ \cite{CMV5}.
It is worth pointing out that, nevertheless, this freedom has nothing to
do with the scaling of the field, leaving intact the time
evolution. If a choice of momentum and of invariant complex structure
permits a unitary implementation of the dynamics, the same complex
structure leads to a unitary evolution for any other admissible
choice of the momentum canonically conjugate to the field. In other
words, this freedom to change the momentum by adding a time dependent
contribution linear in the configuration field variable, when available,
does not allow one to reach a new
representation satisfying our criteria.

As we have explained in the Introduction, a framework
where our results find a natural application is in the quantization of
(inhomogeneous) perturbations around a closed FRW spacetime.
In this context, the simplest system is
a scalar field coupled to a homogeneous
and isotropic universe with compact spatial sections.
This system is specially relevant in {cosmology.} On the one hand,
the considered perturbations provide the seeds for
structure formation. On the other hand, those perturbations
explain the anisotropies imprinted in the power spectrum of
the cosmic microwave background (CMB). Our criteria to eliminate the
quantization ambiguities can now be applied in their quantum
treatment and the subsequent analysis of the power spectrum.

Although the discussion that we have carried out has been focused
on scalar fields, there does not seem to exist any technical or
conceptual obstacle to extend the analysis to other kind of fields,
applying to them our criteria in order to pick up a unique Fock
quantization.
For instance, an interesting case is provided by the traceless and
divergenceless tensor perturbations of the metric around an FRW
spacetime with compact spatial topology. These tensor
perturbations describe gravitational waves.
The primordial gravitational waves generated in the early
universe can also contribute to the power spectrum of the
CMB, in the form of tensor modes. In fact, with a convenient scaling
and in conformal time, these tensor perturbations satisfy again
equations of motion like those for a free field
with a time dependent quadratic potential in a static spacetime
when the perturbations of
the energy-momentum tensor are isotropic
(see Ref. \cite{bardeen}). Let us mention also the
case of fermionic fields. The study of the perturbations around a
closed FRW spacetime produced by fermions of constant mass was
carried out in Ref. \cite{death_hall}, where a quantization was achieved
after expanding the
perturbations in spinor harmonics on the three-sphere.
Preliminary calculations indicate that the kind of techniques
employed here can be extended to deal as well with the uniqueness
of the Fock quantization for fermions. It is worth emphasizing
that the criteria for this uniqueness are the natural implementation
of the symmetries of the field equations and the unitarity of the
evolution. For cases other than the scalar field (and gravitational
waves, as noticed above), these criteria may
not necessarily imply that the selected field description corresponds
to a field propagating in a stationary background.

The Fock quantization of fields in the context of modern approaches
to quantum cosmology is another interesting framework where our
results can have applications. One of the most promising
approaches is what nowadays is called Loop Quantum
Cosmology \cite{lqcboj,lqcash,lqcGAMM}. LQC employs the techniques
of Loop Quantum Gravity (LQG) \cite{lqgthi,lqgrov,lqgashlew}
in the study of models of
interest in cosmology, obtained from General Relativity by the
imposition of certain symmetries. In the specific
case of an FRW spacetime coupled to a scalar field
(see Refs. \cite{aps,acs,mmo}), where homogeneity and isotropy
are imposed, LQC predicts that the classical Big Bang singularity
is replaced by a Big Bounce, which connects the observed branch
of the universe with a previous branch in the evolution.
For semiclassical states with certain properties \cite{taveras},
the evolution is peaked around a trajectory which shows a
behavior different from the classical one in Einstein's theory.
Then, one could use such a trajectory to define an effective,
{\sl quantum corrected} background.
If inhomogeneous matter fields are introduced, their
scaling by background functions would then provide
a different time dependent scaling with
respect to the conventional case in General Relativity.
At the quantum level, the combination of the use of loop
techniques for the homogeneous background with a
standard  Fock quantization of the inhomogeneous fields, which propagate
in it, is known in the literature as {\sl hybrid quantization}
\cite{hyb1,hyb2,hyb3,hyb4}.
This quantization procedure assumes that the most relevant
quantum geometry effects (characteristic of LQG) are those that
affect the homogeneous degrees of freedom of the gravitational
field. A family of systems in which the application of such a
quantization procedure seems most natural is the already commented
case of perturbations around FRW spacetimes. Our
analysis provides a
unique Fock quantization for those perturbations, which ought to be
recovered from LQC (either with an hybrid or with a genuinely loop
quantization) in regimes in which the behavior of the
degrees of freedom of the background can be described
satisfactorily by an effective trajectory.

In conclusion, we have completed the demonstration
that a scalar field with time dependent mass
in an FRW spacetime with compact spatial
sections admits (within the considered infinite family of possibilities
that respect the linear structure of the theory, and up to equivalence)
a unique Fock quantization
where the vacuum is invariant under the symmetries of the field
equation and the dynamical evolution is unitary. In this sense,
it is not only that one does not have to renounce to unitarity
in the context of quantum cosmology, but, furthermore,
the requirement of unitarity has the remarkable counterpart of
selecting a unique Fock description among all those that incorporate
the symmetry invariance of the system.

\section*{Acknowledgements}

This work has been supported by the grants DGAPA-UNAM IN108309-3
from Mexico, MICINN FIS2008-06078-C03-03 and
Consolider-Ingenio Program CPAN (CSD2007-00042) from Spain,
and CERN/FP/109351/2009 from Portugal. J.O. acknowledges
CSIC for the grant No. JAE-Pre\_08\_00791. He thanks I.G.C.
at Penn State for warm hospitality during part of the period of
development of this work.

\appendix
\section{Application of the Lebesgue dominated convergence}

In Sec. \ref{sec:uniq} we used the fact that $\sin^2[(n+1)\tau]$,
or more generally $\sin[(n+1)\tau]\sin[(n+1)\tau-\delta_n]$ (from which
the previous case is recovered by taking $\delta_n=0$), cannot tend to
zero in the limit $n\to\infty$
on any subsequence of the positive integers
$\forall \tau\in	\bar{\mathbb{I}}$.
In this appendix we are going to prove an even more
general result. Let
\begin{equation}
\mathbb{U}=\{ u_n,\ n\in\mathbb{N}^+\}
\end{equation}
be a monotonous and diverging sequence
of positive real numbers; i.e. $u_{n+1}>u_n$ $\forall n\in\mathbb{N}^+$,
with $\{u_n\}$ unbounded. Let also
\begin{equation}
\mathbb{D}=\{\delta_n, n\in\mathbb{N}^+\}\end{equation}
be a sequence of phases [namely, real numbers, identified modulo
$2\pi$], and let $L>0$ be an arbitrary positive number.
Then, the sequences of values
\begin{equation}
x_n(\tau)=\sin(u_n\tau)\sin(u_n\tau-\delta_n)
\end{equation}
cannot tend to zero $\forall \tau\in {[0,L]}$.

To prove this, we assume from the start that the sequence formed by
$\cos(\delta_n)$
with $\delta_n\in\mathbb{D}$ does not tend to zero when $n$ tends to
infinity. We will show below that there is no loss of generality
in making this assumption. A straightforward computation shows that
%%%
%\begin{eqnarray}
%\int_0^Lx_n(\tau)d\tau &=& \cos(\delta_n)\left[\frac{L}{2}-
%\frac{\sin(2u_nL)}{4u_n}\right]-
%\frac{\sin(\delta_n)}{2u_n}\sin^2(u_nL)\nonumber \\
%&=& \frac{L}{2}\cos(\delta_n)-\cos(u_nL-\delta_n)\frac{\sin(u_nL)}{2u_n}.
%\end{eqnarray}
%%
\begin{equation}
\int_0^Lx_n(\tau)d\tau = \frac{L}{2}\cos(\delta_n)-
\cos(u_nL-\delta_n)\frac{\sin(u_nL)}{2u_n}.
\end{equation}
%%%%%%%%%%%%%
Taking into account the range of the trigonometric functions,
we get the following bounds, valid for all positive integers $n$:
\begin{equation}
\label{1bound}
\frac{L}{2} \cos(\delta_n)+\frac{1}{2u_n}\geq\int_0^Lx_n(\tau)d\tau
\geq\frac{L}{2} \cos(\delta_n)-\frac{1}{2u_n}.
\end{equation}
Given that $\cos(\delta_n)$ does not tend to zero, there exists a
subsequence $\mathbb{M}''\subset\mathbb{N}^+$ and a number
$\Delta>0$ such that $|\cos(\delta_n)|\geq\Delta$,
$\forall n\in \mathbb{M}''$. Thus, there exists a subsequence
$\mathbb{M}'\subset \mathbb{M}''$ such that
\begin{equation}\label{case1}
\cos(\delta_n)\geq\Delta\qquad \forall n\in \mathbb{M}',
\end{equation} or
\begin{equation} \cos(\delta_n)\leq -\Delta \qquad \forall n
\in \mathbb{M}'
\end{equation}
(both types of sequences may exist).

Let us consider for the moment the first case \eqref{case1}.
Since the sequence of positive numbers $1/u_n$ (with $u_n\in\mathbb{U}$)
tends to zero for large $n$, one can find a positive integer
$n_0\in \mathbb{M}'$ such that
$L\Delta>1/u_{n_0}$. Moreover, since $u_{n+1}>u_n$ in $\mathbb{U}$,
one gets from the second inequality in Eq. (\ref{1bound}) that the
considered integral is bounded from below
on the given sequence by a positive number:
\begin{equation}
\label{2bound}
\int_0^Lx_n(\tau)d\tau
\geq\frac{L\,\Delta}{2}-\frac{1}{2u_{n_0}}>0,\quad \forall n>n_0,\quad
n\in \mathbb{M}'.
\end{equation}

It is clear that the second possibility, i.e. the existence of a
sequence $\mathbb{M}'$ such that $\cos(\delta_n)\leq -\Delta$,
leads to a negative upper bound by similar arguments.
Taking into account the two possibilities, we conclude that
[assuming that $\cos(\delta_n)$ does not tend to zero]
there exist positive numbers $\Delta$, $n_0\in\mathbb{N}^+$,
and $M=L\Delta-1/u_{n_0}$, as well
as a subsequence $\mathbb{M}\subset \mathbb{N}^+$ such that
\begin{equation}
\label{3bound}
\left|\int_0^Lx_n(\tau)d\tau\right|
\geq \frac{M}{2}\quad \forall  n\in \mathbb{M}.
\end{equation}
The sequence $\mathbb{M}$ is formed by those elements of $\mathbb{M}'$
such that $n>n_0$.

Suppose now that the sequence of
functions $x_n(\tau)$ converges to the zero function on $[0,L]$.
Since the functions $|x_n(\tau)|$ are obviously
bounded from above by the constant unit function, we are in
the conditions of Lebesgue dominated
convergence \cite{Vonneu},
and it follows that the sequence of integrals
$\int_0^Lx_n(\tau)d\tau$ must converge to the integral of
the zero function, i.e. to zero.
But this conclusion is obviously contradicted by the bound
obtained in Eq. (\ref{3bound}).
Therefore, it is not possible that the values of $x_n(\tau)$
converge to zero $\forall\tau\in [0,L]$.

To conclude the proof, it only remains to consider the situation
in which $\cos(\delta_n)$ tends to zero for large $n$.
In that case, the vanishing limit of $x_n(\tau)$
implies that $\sin(2u_n\tau)$ must tend to zero,
and therefore so must $\sin^2(2u_n\tau)$. But this last
situation is covered by the proof presented above. It suffices
to make all the phases $\delta_n$ identically null,
and identify the sequence $\{2u_n\}$ as the new sequence $\mathbb{U}$,
since the positive real numbers $2u_n$ form a monotonous and diverging
sequence if so do the $u_n$'s.

\bibliographystyle{plain}

\begin{thebibliography}{99}

\bibitem{Vonneu} B. Simon, {\it Topics in Functional
Analysis}, edited by R.F. Streater (Academic Press, London,
England, 1972).

\bibitem{wald} R.M. Wald, {\it Quantum Field Theory in
Curved Spacetime and Black Hole Thermodynamics} (Chicago
University Press, Chicago, 1994).

\bibitem{a-m} A. Ashtekar and A. Magnon, Proc. R. Soc. Lond.
A {\bf 346}, 375 (1975); A. Ashtekar and A. Magnon-Ashtekar,
Pramana {\bf 15}, 107 (1980).

\bibitem{jackiw} R. Floreanini, C.T. Hill, and R. Jackiw.
Ann. Phys. {\bf 175}, 345 (1987).

\bibitem{gowdy} R.H. Gowdy, Phys. Rev. Lett. {\bf 27},
826 (1971); Ann. Phys. (N.Y.) {\bf 83}, 203
(1974).

\bibitem{unit-gt3} A. Corichi, J. Cortez, and G.A. Mena
Marug\'an, Phys. Rev. D {\bf 73}, 041502 (2006); Phys. Rev.
D {\bf 73}, 084020 (2006).
%%CITATION = PHRVA,D73,084020;%%
%%CITATION = PHRVA,D73,041502;%%



\bibitem{feasab} J. Cortez and G.A. Mena Marug\'an,
Phys. Rev. D \textbf{72}, 064020 (2005).
%%CITATION = PHRVA,D72,064020;%%

\bibitem{unique-gowdy-1} A. Corichi, J. Cortez, G.A. Mena
Marug\'an, and J.M. Velhinho, Classical Quantum Gravity {\bf 23},
6301 (2006); Phys. Rev. D {\bf 76}, 124031 (2007).
%%CITATION = CQGRD,23,6301;%%
%%CITATION = PHRVA,D76,124031;%%

\bibitem{CMV5} J. Cortez, G.A. Mena Marug\'an, and J.M.
Velhinho, Phys. Rev. D {\bf 75}, 084027 (2007).
%%CITATION = PHRVA,D75,084027;%%

\bibitem{cmsv} J. Cortez, G.A. Mena Marug\'an, R. Ser\^odio,
and J.M. Velhinho, Phys. Rev. D {\bf 79}, 084040 (2009);
J. Cortez, G.A. Mena Marug\'an, and J.M.
Velhinho, Classical Quantum Gravity {\bf 25}, 105005 (2008).
%%CITATION = CQGRD,25,105005;%%
%%CITATION = PHRVA,D79,084040;%%

\bibitem{CMV8} J. Cortez, G.A. Mena Marug\'an, and J.M.
Velhinho, Phys. Rev. D {\bf 81}, 044037 (2010).
%%CITATION = PHRVA,D81,044037;%%

\bibitem{CMOV} J. Cortez, G.A. Mena Marug\'an, J. Olmedo, and
J.M. Velhinho, J. Cosmol. Astropart. Phys. 10 (2010) 030.
%%CITATION = JCAPA,1010,030;%%



\bibitem{lif} E. Lifshitz, Zh. Eksp. Teor. Fiz. {\bf
16}, 587 (1946); E. Lifshitz and I.M. Khalatnikov,
Adv. Phys. {\bf 12}, 185 (1963).

\bibitem{bardeen} J.M. Bardeen, Phys. Rev. D {\bf 22}, 1882
(1983).

\bibitem{bar2} R. Bradenberger, R. Kahn, and W.H.
Press, Phys. Rev. D {\bf 28}, 1809 (1983).

\bibitem{mukhanov} V. Mukhanov, {\it{Physical Foundations of
Cosmology}} (Cambridge University Press, Cambridge,
England, 2005).

\bibitem{mukhanov1} V.F. Mukhanov, H.A. Feldman, and R.H.
Bradenberger, Phys. Rep. 215, 203 (1992).

\bibitem{hall_haw} J.J. Halliwell and S.W. Hawking, Phys.
Rev. D {\bf 31}, 1777 (1985).

\bibitem{bkl} J. Bi\u{c}\'ak, J. Katz, and D. Lynden-Bell, 
Phys. Rev. D {\bf 76}, 063501 (2007).

\bibitem{quad1} G. Efstathiou, Mon. Not. R. Astron. Soc.
{\bf 343}, L95 (2003).

\bibitem{quad2}  A. Lasenby and C. Doran, Phys.
Rev. D {\bf 71}, 063502 (2005).

\bibitem{CCAP} A. Corichi, J. Cortez, and H. Quevedo,
Ann. Phys. (N.Y.) {\bf 313}, {446} (2004).
%%CITATION = APNYA,313,446;%%

\bibitem{SH} R. Honegger and A. Rieckers,
J. Math. Phys. {\bf 37}, 4292 (1996);
D. Shale, Trans. Am. Math. Soc. {\bf 103}, 149 (1962).

\bibitem{GS} U.H. Gerlach and U.K. Sengupta, Phys.
Rev. D {\bf 18}, 1773 (1978).

\bibitem{jantzen} R.T. Jantzen, J. Math. Phys. {\bf
19}, 1163 (1978).

\bibitem{schur} A.A. Kirillov, {\it Elements of the
Theory of Representations} (Springer-Verlag, New York,
1976).

\bibitem{MTV} J.M. Mour\~ao, T. Thiemann, and J.M. Velhinho,
J. Math. Phys. {\bf 40}, 2337 (1999).
%%CITATION = JMAPA,40,2337;%%

\bibitem{death_hall} P.D. D{'}Eath and J.J. Halliwell, Phys Rev.
D {\bf 35}, 1100 (1987).

\bibitem{lqcboj} M. Bojowald, Living Rev.
Relativity {\bf 11}, 4 (2008).

\bibitem{lqcash} A. Ashtekar, Nuovo Cimento
{\bf B122}, 135 (2007).

\bibitem{lqcGAMM} G.A. Mena Marug\'an,
AIP Conf. Proc. {\bf 1130}, 89 (2009).
%%CITATION = APCPC,1130,89;%%

\bibitem{lqgthi} T. Thiemann, {\it Modern Canonical Quantum
General Relativity} (Cambridge University Press, Cambridge,
England, 2007).

\bibitem{lqgrov} C. Rovelli, {\it{Quantum Gravity}}
(Cambridge University Press, Cambridge, England, 2004).

\bibitem{lqgashlew} A. Ashtekar and J. Lewandowski,
Classical Quantum Gravity {\bf 21}, R53 (2004).

\bibitem{aps} A. Ashtekar, T. Paw{\l}owski, and P. Singh,
Phys. Rev. Lett. {\bf 96}, 141301 (2006); Phys. Rev. D {\bf
73}, 124038 (2006); Phys. Rev. D {\bf 74}, 84003 (2006).

\bibitem{acs}A. Ashtekar, A. Corichi, and P. Singh,
Phys. Rev. D {\bf 77}, 024046 (2008).

\bibitem{mmo} M. Mart\'in-Benito, G.A. Mena Marug\'an, and J.
Olmedo, Phys. Rev. D {\bf 80}, 104015 (2009).
%%CITATION = PHRVA,D80,104015;%%

\bibitem{taveras} V. Taveras, Phys. Rev. D {\bf 78},
064072 (2008).

\bibitem{hyb1} M. Mart\'in-Benito, L.J. Garay, and G.A. Mena
Marug\'an, Phys. Rev. D {\bf 78}, 083516 (2008).
%%CITATION = PHRVA,D78,083516;%%

\bibitem{hyb2} G.A. Mena Marug\'an and M. Mart\'in-Benito,
Int. J. Mod. Phys. A {\bf 24}, 2820 (2009).
 %%CITATION = IMPAE,A24,2820;%%


\bibitem{hyb3} L.J. Garay, M. Mart\'in-Benito, and G.A. Mena
Marug\'an, Phys. Rev. D {\bf 82}, 044048 (2010).
%%CITATION = PHRVA,D82,044048;%%

\bibitem{hyb4} M. Mart\'in-Benito, G.A. Mena Marug\'an, and E.
Wilson-Ewing, Phys. Rev. D {\bf 82}, 084012 (2010).
 %%CITATION = ARXIV:1006.2369;%%

\end{thebibliography}

\end{document}